\documentstyle[ApJ,times,PSfig]{article}

\begin{document}

\def\simg{\mathrel{%
      \rlap{\raise 0.511ex \hbox{$>$}}{\lower 0.511ex \hbox{$\sim$}}}}
\def\siml{\mathrel{%
      \rlap{\raise 0.511ex \hbox{$<$}}{\lower 0.511ex \hbox{$\sim$}}}}
\def\Mesz{M\'esz\'aros~}
\def\ie{i.e$.$~} \def\eg{e.g$.$~} \def\etal{et al$.$~} 
\def\eq{eq$.$~} \def\eqs{eqs$.$~} \def\deg{^{\rm o}} \def\dd{{\rm d}}
\def\beq{\begin{equation}} \def\eeq{\end{equation}}
\def\epsel{\varepsilon_e} \def\epse1{\varepsilon_{e,-1}} 
\def\epsmag{\varepsilon_B} \def\eBzero{\tilde{\varepsilon_B}} 
\def\eBone{\varepsilon_{B,-1}} \def\eBtwo{\varepsilon_{B,-2}} 
\def\eBthree{\varepsilon_{B,-3}} \def\eBfour{\varepsilon_{B,-4}}
\def\E53{E_{53}} \def\nex0{n_{*,0}} \def\Gm02{\Gamma_{0,2}} 
\def\D28{D_{28}^{-2}} \def\nuo{\nu_{14.6}}
\def\Tdone{T_{d,-1}} \def\Tdtwo{T_{d,-2}} \def\Tdthree{T_{d,-3}}
 
\title{Analytic Light-Curves of Gamma-Ray Burst Afterglows: \\
       Homogeneous versus Wind External Media}

\author{A. Panaitescu}
\affil{Dept of Astrophysical Sciences, Princeton University, Princeton, NJ 08544}
\centerline{and}
\author{P. Kumar}
\affil{School of Natural Sciences, Institute for Advanced Study, Princeton, NJ 08540}

\begin{abstract}

 Assuming an adiabatic evolution of a Gamma-Ray Burst (GRB) remnant interacting 
 with an external medium, we calculate the injection, cooling, and absorption
 break frequencies, and the afterglow flux for plausible orderings of the
 break and observing frequencies. The analytical calculations are restricted 
 to a relativistic remnant and, in the case of collimated ejecta, to the phase 
 where there is an insignificant lateral expansion. Results are given for both a
 homogeneous external medium and for a wind ejected by the GRB progenitor.

 We compare the afterglow emission at different observing frequencies, for each 
 type of external medium. It is found that observations at sub-millimeter frequencies 
 during the first day provide the best way of discriminating between the two 
 models. By taking into account the effect of inverse Compton (IC) scatterings on the 
 electron cooling, a new possible time-dependence of the cooling break is identified. 
 The signature of the up-scattering losses could be seen in the optical synchrotron 
 emission from a GRB remnant interacting with a pre-ejected wind, as a temporary mild 
 flattening of the afterglow decay. The up-scattered radiation itself should be 
 detected in the soft X-ray emission from GRB remnants running into denser external 
 media, starting few hours after the main event. 

\end{abstract}


\section{Introduction}

 One of the most important issues regarding Gamma-Ray Bursts (GRBs) is the nature 
of the object that releases the relativistic ejecta generating the high energy emission 
of the main event and the lower frequency emission during the ensuing afterglow. 
Some insight about the GRB progenitor can be obtained from the properties of the 
circum-burst medium, which can be inferred from the features of the afterglow emission. 
If the ejecta is expelled during the merging of two compact objects (\Mesz \& Rees 
1997b), it is expected that the medium surrounding the GRB source is homogeneous. 
However, if a collapsing massive star (Woosley 1993, Paczy\'nski 1998) is the origin 
of the relativistic fireball, the circum-burst medium is the wind ejected by the star 
prior to its collapse, whose density decreases outwards. The two models differ in the 
dependence on radius of the particle density of the circum-burst medium which the GRB 
remnant interacts with, and in the value of this density at the deceleration length-scale. 
The former modifies the rate of decline of the afterglow, while the latter determines the 
overall afterglow brightness. Therefore it is possible to correlate afterglow emission 
features to a specific type of external medium.

 Significant work in this direction has been done by many researchers. 
The two afterglows that exhibited breaks consistent with the effects arising from 
strong collimation of ejecta -- GRB 990123 (Kulkarni \etal 1999a) and GRB 990510 
(Stanek \etal 1999, Harrison \etal 1999) -- indicate that the external gas was 
homogeneous (recent work by Kumar \& Panaitescu 2000 shows that jets interacting 
with winds cannot produce sharp breaks in the afterglow light-curve).
The optical emission of three afterglows had a steeper than usual decline.
GRB 970228 decayed as $T^{-1.7}$ after the subtraction of an underlying supernova 
emission (Reichart 1999, Galama \etal 2000). The light-curve of GRB 980326 fell off
as $T^{-2.1}$ (Groot \etal 1998) and an emission in excess of the early time extrapolation 
was detected $\sim 20$ days after the main event, indicating a supernova contribution
(Bloom \etal 1999). A $T^{-2}$ decay  was observed for the afterglow of GRB 980519 
(Halpern \etal 1999). Such steep declines can be produced either by a fireball 
interacting with a pre-ejected wind (Chevalier \& Li 1999) and an electron 
index around 3, or by a narrow jet expanding laterally in a homogeneous external 
medium and an electron index slightly larger than 2. 
Chevalier \& Li (1999) found that the radio emission of the afterglow of GRB 980519 
is consistent with an external wind; however Frail \etal (2000) point out that the
interstellar scintillation present in the radio data does not allow ruling out 
the jet model. Nevertheless, the existence of supernovae associated with GRB 970228 
and GRB 980326 points toward a massive star as the origin of these bursts, implying 
a pre-ejected wind as the external medium.
From the analysis of the optical radio emission of the afterglow of GRB 970508, 
Chevalier \& Li (2000) conclude that the surrounding medium was a wind. 
Frail, Waxman \& Kulkarni (2000) argue that the same radio afterglow can be explained 
by a homogeneous external medium.

 In this work we investigate the differences between the light-curves of afterglows
arising for each type of external medium, with the aim of finding ways for distinguishing
between the two models. This study is done within the usual framework of a relativistic 
remnant interacting with a cold external gas. As the fireball is decelerated, a shock 
front sweeps up the external gas, accelerating relativistic electrons and generating a 
magnetic field in the shocked gas. We ignore the emission from electrons accelerated
by the reverse shock which propagates through the ejecta at very early times. At optical 
wavelengths this emission is short lived, lasting up to few tens of seconds after the 
main event (Sari \& Piran 1999), but it could be important for the radio emission until 
few days (Kulkarni \etal 1999b). 

 Analytical afterglow light-curves for spherical remnants interacting with homogeneous 
external media have been previously published by Sari, Piran \& Narayan (1998).  
Features of afterglows from spherical fireballs, such as peak flux, break frequencies, 
and time evolution of fluxes at a fixed frequency, have been studied by \Mesz \& Rees 
(1997a), Waxman (1997b), Wijers \& Galama (1999), and Dai \& Lu (2000) for homogeneous 
media, and by Chevalier \& Li (2000) for pre-ejected winds. In this work we present 
and compare analytical and numerical light-curves at various observing frequencies, 
covering all cases of interest, for both types of external media, taking into account 
the differential arrival-time delay and Doppler boosting due to the spherical shape of
the source. We take into account first order IC scattering, calculate its effect on 
the electron cooling and on the afterglow synchrotron emission, and study briefly the 
high-energy emission resulting from the up-scattering of synchrotron photons. The possible 
importance of IC scatterings for the early afterglow emission was pointed out by Waxman 
(1997a) and Wei \& Lu (1998).

\vspace*{5mm}
\section{Simple Dynamics of Relativistic Remnants}

 For the calculation of the afterglow emission it is necessary to know how the
remnant Lorentz factor $\Gamma$ evolves with observer time $T$, as all other
quantities that appear in the expression of the spectral flux are functions
of $\Gamma$ and of the remnant radius $r$ and external medium density $n(r)$.
We shall assume that the remnant is adiabatic, \ie the energy carried away by the
emitted photons is a negligible fraction of the total energy of the remnant.
This assumption is correct if the energy density of the electrons accelerated 
at the shock front is a fraction $\epsel \ll 1$ of the total energy density in
post-shock fluid or if most of the electrons are adiabatic, \ie their radiative
cooling timescale exceeds that of the adiabatic losses due to the remnant expansion.

 Assuming that the internal energy of the ejecta is negligible compared to its
rest-mass energy and that the ratio internal-to-rest mass energy in the energized
external medium is $\Gamma - 1$ (\ie its ``temperature" tracks that of the freshly
shocked gas), conservation of energy leads to
\beq
 m(r)\Gamma^2 + M_{fb}\Gamma - [m(r)+M_{fb}\Gamma_0] = 0
\label{cons}
\eeq 
where $M_{fb}$ and $\Gamma_0$ are the initial mass and Lorentz factor of 
the fireball (whose energy is $E=M_{fb}\Gamma_0$) and
\beq
 m(r) = \frac{4\pi}{3-s} m_p n(r) r^3 
\label{mr}
\eeq
is the mass of swept-up material ($m_p$ being the proton's mass). The external
medium particle density is
\beq
 n(r) = A r^{-s} \;,
\label{nr}
\eeq
with $s=0$ for a homogeneous medium and $s=2$ for a wind ejected by the GRB 
progenitor at a constant speed.

 Equation (\ref{mr}) is valid if the remnant is spherical, but can also be used
for collimated ejecta when the lateral spreading (Rhoads 1999) is insignificant
if the quantity $E$ above is defined as the energy the fireball would have if it
were spherical. Throughout this work we shall assume that the remnant is a jet 
with an initial half-angle larger than $\simg 20\deg$, in which case the sideways 
expansion is negligible during the relativistic phase. 
The following analytical calculations of the afterglow emission can be extended 
to sideways expanding jets and non-relativistic remnants by first determining 
$\Gamma (r)$. The set of coupled differential equations describing the evolution 
of the jet Lorentz factor and its opening can be solved analytically for $s=0$ 
(Rhoads 1999). The lack of a good approximation for the jet dynamics in the case 
of pre-ejected winds is the main motivation for restricting the following analytical 
calculations to spherical or wide-angle remnants.

 The solution of equation (\ref{cons}) is
\beq
 \Gamma(r) = \frac{1}{2} \left[ \sqrt{4x^{3-s} + 1 + (2x^{3-s}/\Gamma_0)^2} - 1 
                        \right] x^{s-3} \Gamma_0 \;,
\label{Gm}
\eeq 
where $x$ is the radial coordinate $r$ scaled to 
\beq
 r_0 = \left( \frac{3-s}{4\pi} \frac{E}{m_p c^2 A \Gamma_0^2} \right)^{1/(3-s)} \;,
\label{r0}
\eeq
the deceleration length-scale, at which $m(r_0) = E/(c^2 \Gamma_0^2) = M_{fb}/\Gamma_0$.
The result given in equation (\ref{Gm}) is also valid in the non-relativistic regime. 
For $x \ll 1\;$ $\Gamma \siml \Gamma_0$, while for $1 \ll x \ll x_{nr}$ we  find
$\Gamma = x^{-(3-s)/2} \Gamma_0$. Here $x_{nr}=(\Gamma_0^2/3)^{1/(3-s)}$ marks the 
end of the relativistic regime: $\Gamma(x_{nr})=2$. For the ease of analytical calculations 
we shall assume that the power-law behavior of $\Gamma$ lasts from $x=1$ to $x=x_{nr}$.

 The Lorentz factor given in equation (\ref{Gm}) represents a ``dynamical'' average
of the Lorentz factors at which different regions of the shocked remnant move
(the Blandford -- McKee solution). The Lorentz factor of the shock front that 
propagates into the external gas $\Gamma_{sh} = \sqrt{2}\, \Gamma$, with $\Gamma$ given 
by equation (\ref{Gm}), matches that given in equation (69) of Blandford \& McKee 
(1976) for the power-law regime $1 \ll x \ll x_{nr}$ if $E$ is multiplied by 
$(17-4s)/(12-4s)$. This correction factor ($17/8$ for $s=0$ and $9/4$ for $s=2$) 
will be used in the following results.

 The constant $A$ in equation (\ref{nr}) is the number density $n_*$ of the
external homogeneous medium for $s=0$, while for $s=2$
\beq
 A = \frac{1}{4\pi} \frac{\stackrel{.}{M}}{m_p v} 
   = 3.0 \times 10^{35}\; A_* \;\; {\rm cm^{-1}} \;,
\label{A}
\eeq
where $\stackrel{.}{M}$ is the mass loss rate of the massive star that
ejected the wind at constant speed $v$, and $A$ was scaled to
\beq
 A_* = \frac{ \stackrel{.}{M}/10^{-5}\,{\rm M_{\odot} yr^{-1}} }
            { v/10^3\,{\rm km\, s^{-1}} } \;,
\label{Astar}
\eeq
as in Chevalier \& Li (2000) for a Wolf-Rayet star. For further calculations it is 
convenient to use equation (\ref{nr}) in the form $n(r) = n_0 (r/r_0)^{-s}$ where 
$n_0 = n_*$ for $s=0$, while for $s=2$ 
\beq
 n_0 = 1.9 \times 10^4\; \E53^{-2} \Gm02^4 A_*^3 \;\; {\rm cm^{-3}}
\label{n0}
\eeq
is the wind particle density at the deceleration radius (\eq [\ref{r0}]):
\beq 
 (s=2) \qquad  r_0 = 4.0 \times 10^{15}\; \E53 \Gm02^{-2} A_*^{-1} \;\; {\rm cm} \;.
\label{r02} 
\eeq
The usual notation $C_n=10^{-n} C$ is used throughout this work.

 Note that, for the reference values used here, the deceleration radius in the wind 
model is 1.5 orders of magnitude lower than that for a homogeneous external medium:
\beq
 (s=0) \qquad r_0 = 1.3 \times 10^{17}\; \E53^{1/3} \Gm02^{-2/3} \nex0^{-1/3} 
               \;\; {\rm cm} \;,
\label{r00}
\eeq 
and even smaller for higher initial fireball Lorentz factors or slower
winds (\ie a larger parameter $A_*$). If GRBs are due to internal shocks
occurring in unstable relativistic fireballs (Rees \& \Mesz 1994, Paczy\'nski 
\& Xu 1994, Piran 1999), then the external shock resulting from the interaction of the 
fireball with the pre-ejected (non-relativistic) wind may occur before the 
internal shocks are over. In this case successive internal collisions occur
when faster parts of the ejecta catch up with the decelerating leading
edge of the fireball, a scenario suggested within the $s=0$ model by
Fenimore \& Ramirez-Ruiz (2000), but which is more likely to happen
if the external medium is the gas ejected by a massive star. The GRB
itself would then exhibit the erratic variability characteristic of
internal shocks until a time of the order
\beq
 \frac{r_0}{c \Gamma_0^2} \simeq 10\,\E53 \Gm02^{-4} A_*^{-1} \;\; s \;,
\eeq
(which has a strong dependence on $\Gamma_0$), after which there may be
significant emission from internal shocks on the outermost part of the 
fireball and from the external shock that plows through the external gas. 
The former mechanism generates pulses of increasing duration as the fireball 
expands, while the later leads to a continuous emission. 
 
 The time $T$ when the observer receives a photon emitted along the line 
of sight toward the fireball center can be calculated by integrating
\beq
 \dd T = (1-\beta) \dd t = \frac{1}{2} \frac{\dd t}{\Gamma^2} \;, 
\label{dT}
\eeq
where $\beta$ is the shocked fluid speed and $t = r/c$ is the time measured 
in the laboratory frame. Approximating the solution given in equation (\ref{Gm})
with $\Gamma = \Gamma_0$ for $x < 1$ and $\Gamma = x^{-(3-s)/2} \Gamma_0$ for
$1 < x < x_{nr}$, one obtains
\beq
 T=T_0 (x^{4-s}+3-s)\;, \quad T_0\equiv \frac{1}{2(4-s)} \frac{r_0}{c\Gamma_0^2} \;.
\label{Time}
\eeq
Note that $T$ given in equation (\ref{Time}) is the earliest time a photon emitted
by the remnant at time $t$ can reach the observer. Photons emitted by the fluid 
moving at an angle $\theta=1/\Gamma$ off the center--observe axis arrive at 
$T=t(1-\cos \theta)=  t/2\Gamma^2$, which is a factor $4-s$ larger than 
the time corresponding to $\theta=0$: $T= t/2(4-s)\Gamma^2$. 

 From equation (\ref{Time}) $r(T)$ can be found and then substituted in the 
expressions for $\Gamma(r)$ and $n(r)$ to obtain these quantities as a function of
the observer time. For the power-law phase the results are 
\beq
 (s=0)\qquad \qquad \Gamma (T)= 6.3\; \E53^{1/8} \nex0^{-1/8} T_d^{-3/8} \;,
\label{Gms0}
\eeq
\beq
 (s=0)\qquad r(T)= 8.2\times 10^{17}\; \E53^{1/4} \nex0^{-1/4} T_d^{1/4} \;{\rm cm}\;,
\label{rs0}
\eeq
$n=n_*$ (constant) for a homogeneous external medium and
\beq
 (s=2) \qquad \qquad \Gamma (T) = 7.9\; \E53^{1/4} A_*^{-1/4} T_d^{-1/4} \;,
\label{Gms2}
\eeq
\beq
 (s=2)\qquad r(T) = 6.4 \times 10^{17}\; \E53^{1/2} A_*^{-1/2} T_d^{1/2} \;{\rm cm}\;,
\label{rs2}
\eeq
\beq
 (s=2)\qquad \qquad n(T) = 0.73\; \E53^{-1} A_*^2 T_d^{-1} \;{\rm cm^{-3}}\;,
\label{ns2}
\eeq
for an external wind, $T_d$ being the observer time measured in days. Note that,
at least for the scaling values chosen here, $\Gamma$, $r$, and $n$ have about the 
same values at $T=1$ day in both models. Also note that the above quantities (and 
thus the afterglow emission) are independent of the fireball initial Lorentz factor 
$\Gamma_0$.

\section{Break Frequencies}

 Within the synchrotron emission model there are three expected breaks in the afterglow
spectrum: $(i)$ an injection break, at the synchrotron frequency $\nu_i$ at which the 
bulk of the electrons injected by the shock front radiate, $(ii)$ a cooling break, at 
the synchrotron frequency $\nu_c$ of electrons whose radiative cooling time equals the 
expansion timescale, and $(iii)$ an absorption break, at $\nu_a$ below which the 
synchrotron photons are absorbed by electrons in free-free transitions in a magnetic 
field (synchrotron self-absorption).

 The break frequencies can be calculated if the distribution of the injected electrons 
and the strength of the magnetic field are known. The distribution of the injected electrons 
is assumed to be ${\cal N}_i (\gamma) \propto \gamma^{-p}$ starting from a minimum random 
Lorentz factor given by
\beq
 \gamma_i = \frac{m_p}{m_e} \epsel (\Gamma - 1)\;,
\label{gmin}
\eeq
where $m_e$ is the electron mass. The energy carried by this electron distribution is a
fraction $\frac{p-1}{p-2}\epsel$ of the total internal energy. The post-shock magnetic field 
strength in the co-moving frame is given by
\beq
 \frac{B^2}{8\pi} = \epsmag m_p c^2 n'_e (\Gamma-1) = 
    4 \epsmag m_p c^2 n(r) (\Gamma-1) \left(\Gamma+\frac{3}{4}\right) \;,
\label{Bsq}
\eeq
where $\epsmag$ is the fractional energy carried by the magnetic field and $n'_e$ is the
co-moving frame electron density behind the shock front.
Equations (\ref{gmin}) and (\ref{Bsq}) are based on that the internal to rest-mass energy 
density ratio in the shocked fluid is $\Gamma-1$; the derivation of the latter equation also 
used that the co-moving particle density is $4\Gamma+3$ times larger than that ahead of the 
shock.

\subsection{Injection Break}

 Using the relativistic Doppler factor $2\Gamma$ corresponding to the motion of 
the source toward the observer (\ie $\theta=0$), the synchrotron emission from a 
power-law distribution of electrons peaks at the observer frame frequency 
\beq
 \nu_i = \frac{3 x_p}{2\pi} \frac{e}{m_e c} \gamma_i^2 B \Gamma 
       = 8.4 \times 10^6\; x_p \gamma_i^2 B \Gamma \; {\rm Hz}\;,
\label{nusy}
\eeq
where the factor $x_p$ is calculated in Wijers \& Galama (1999) for various values 
of the electron index $p$. We shall use $x_p=0.52$, which is strictly correct only 
for $p=2.5$. With the aid of equations (\ref{Gms0}), (\ref{Gms2}), (\ref{ns2}), 
(\ref{gmin}), and (\ref{Bsq}) one obtains:
\beq (s=0) \qquad 
 \nu_i=0.92\times 10^{13}\; \E53^{1/2} \epse1^2 \eBtwo^{1/2} T_d^{-3/2} \; {\rm Hz} \;,
\label{nusy0}
\eeq
\beq (s=2) \qquad 
 \nu_i=1.9\times 10^{13}\; \E53^{1/2} \epse1^2 \eBtwo^{1/2} T_d^{-3/2} \; {\rm Hz} \;.
\label{nusy2}
\eeq
Note that $\nu_i$ has the same scalings with the model parameters for $s=0$
and $s=2$. The ratio of the two frequencies is $\sqrt{\frac{17}{72}}$.

\subsection{Cooling Break}

 The relativistic electrons cool radiatively through synchrotron emission and
IC scatterings of the synchrotron photons on a co-moving frame timescale
\beq
 t'_{rad}(\gamma) = \frac{t'_{sy}(\gamma)}{Y+1} = \frac{6\pi}{Y+1} 
                  \frac{m_e c}{\sigma_e} \frac{1}{\gamma B^2} \;,
\label{trad}
\eeq 
where $t'_{sy}(\gamma)$ is the co-moving frame synchrotron cooling timescale of electrons 
of Lorentz factor $\gamma$, $Y$ is the Compton parameter, and $\sigma_e$ the cross-section 
for electron scatterings\footnote{ 
   The up-scattering of the $\nu_i$ synchrotron photons on the $\gamma_c$- and 
   $\gamma_i$-electrons occurs in the Thomson regime for $T > 10^{-3}$ day, \ie the scattering 
   cross-section in equation (\ref{trad}) is not reduced by the Klein-Nishina effect}. 
Using equation (\ref{Bsq}), the Lorentz factor of the electrons that cool radiatively on a 
timescale equal to the remnant age 
\beq
 t' = \frac{1}{c} \int \frac{\dd r}{\Gamma} = \frac{2}{5-s} \frac{r}{c \Gamma}
\label{tcom}
\eeq
can be written as 
\beq
 \gamma_c = \frac{3\pi(5-s)}{Y+1} \frac{m_e c^2}{\sigma_e} \frac{\Gamma}{B^2 r} =
            \frac{77(5-s)}{(Y+1)\epsmag} \frac{1}{n\Gamma r_{18}} \;,
\label{gc}
\eeq
with $n$ in ${\rm cm^{-3}}$. 

 The observer-frame frequency $\nu_c$ of the cooling break is
\beq
 (s=0) \quad \nu_c = 3.7 \times 10^{14}\; \E53^{-1/2} \nex0^{-1} (Y+1)^{-2}
                         \eBtwo^{-3/2} T_d^{-1/2} \; {\rm Hz} \;,
\label{nucr0}
\eeq
\beq
 (s=2) \quad \nu_c = 3.5 \times 10^{14}\; \E53^{1/2} A_*^{-2} (Y+1)^{-2}
                         \eBtwo^{-3/2} T_d^{1/2} \; {\rm Hz} \;.
\label{nucr2}
\eeq

 Hereafter we shall use the terminology ``radiative electrons" for the case where the 
$\gamma_i$-electrons cool mostly through emission of radiation (\ie $t'_{rad}(\gamma_i)
< t'$ and $\gamma_c < \gamma_i$), and we shall refer to ``adiabatic electrons'' if the 
$\gamma_i$-electrons cool mostly adiabatically (\ie $t'_{rad}(\gamma_i) > t'$ and
$\gamma_i < \gamma_c$).

\subsubsection{Compton Parameter and Electron Distribution}

 For the calculation of the Compton parameter $Y$, we take into account only one 
up-scattering of the synchrotron photons. Multiple IC scatterings of the same photon have 
an important effect on the electron cooling only if the $Y$ parameter for single scatterings 
is above unity. As shown in \S\ref{radels} and \S\ref{adbels} this occurs if 
$1)$ $\epsmag \siml 10^{-2} \epse1$ and if $2)$ $T < T_y$, where $T_y$ is the time when $Y$ 
falls below unity. 

 Using equations (\ref{gmin}) and (\ref{Bsq}), it can be shown that IC scatterings of order
higher than two are suppressed by the Klein-Nishina effect. For adiabatic electrons, a second 
IC scattering occurs in the Thomson regime if 
$3a)$ $T \simg 1\; \E53^{1/3} \nex0^{-1/9} \epse1^{20/9} \eBfour^{2/9} \;{\rm d}$ for $s=0$, 
and if $3b)$ $T \simg 1\; \E53^{1/2} A_*^{-1/4} \epse1^{5/2} \eBfour^{1/4} \;{\rm d}$ for $s=2$. 
If a second IC scattering is ignored in these cases then a higher energy component peaking 
around 1 GeV is left out, otherwise the synchrotron and first IC emissions remain unaltered 
{\sl if} the electrons are adiabatic. 

 The effect of second order up-scatterings is more important when electrons are radiative, 
\ie for $4)$ $T < T_r$ with $T_r$ calculated in \S\ref{radels}, as in this case it reduces 
the intensity of the synchrotron and first stage IC components. With the aid of equations 
(\ref{gmin}) and (\ref{Bsq}) it can be shown that for $s=0$ a second up-scattering occurs 
in the Thomson regime and at $T > 10^{-2}\, {\rm d}$ if 
$5a)$ $n \simg 25\; \E53^{-7/11} \epse1^{-10/11} \eBthree^{-2/11}\; {\rm cm^{-3}}$, 
while for $s=2$ the second IC emission is not suppressed by the Klein-Nishina effect if 
$5b)$ $T \simg 0.05\; \E53^{-1/3} A_*^{11/6} \epse1^{5/6} \eBthree^{2/3} \;{\rm d}$. 

 Concluding, second stage up-scatterings can be ignored if the set of conditions $1),2),3)$  
or $1),4),5)$ are not simultaneously satisfied. For a single up-scattering, the Compton parameter 
is
\beq
 Y = \frac{4}{3} \int_{{\cal N}_e} \gamma^2 \dd \tau_e = 
     \frac{4}{3} \tau_e \int_{{\cal N}_e} \gamma^2 {{\cal N}_e}(\gamma) \dd \gamma \;,
\label{YC}
\eeq
where ${{\cal N}_e}(\gamma)$ is the normalized electron distribution and $\tau_e$ is the 
optical thickness to electron scattering, given by
\beq
 \tau_e=\frac{1}{4\pi} \frac{\sigma_e m(r)}{m_p r^2}=\frac{1}{3-s} \sigma_e nr \;.
\label{tau}
\eeq

 If the injected $\gamma_i$-electrons cool faster than the timescale of their injection,
then $\gamma_c$ given by equation (\ref{gc}) is the typical electron Lorentz factor in 
the remnant, and the electron distribution in the shocked fluid can be approximated by
\beq
 {\cal N}_e^{(r)} \propto  \left\{ \begin{array}{ll} 
           \gamma^{-2} & \gamma_c < \gamma < \gamma_i \\
           \gamma^{-(p+1)} & \gamma_i < \gamma \end{array} \right.  \;,
\label{Ngr}
\eeq
with $p > 2$. In the opposite case most electrons have a random Lorentz factor $\gamma_i$, 
and the electron distribution is
\beq
 {\cal N}_e^{(a)} \propto  \left\{ \begin{array}{ll} 
           \gamma^{-p} & \gamma_i < \gamma < \gamma_c \\
           \gamma^{-(p+1)} & \gamma_c < \gamma \end{array} \right.  \;.
\label{Nga}
\eeq

\subsubsection{Radiative Electrons}
\label{radels}

 For $\gamma_c \ll \gamma_i$ equations (\ref{YC}) and (\ref{Ngr}) lead to
\beq
 Y_r = \frac{4}{3} \gamma_i \gamma_c \tau_e \;.
\label{yr}
\eeq
Substituting $\gamma_c$ with the aid of equation (\ref{gc}), one obtains
\beq
 Y_r (Y_r+1) = \frac{5-s}{8(3-s)} \frac{n'_e m_e c^2 \gamma_i}{B^2/8 \pi} =  
               \frac{5-s}{8(3-s)}\frac{\epsel}{\epsmag} \;,
\label{yr1}
\eeq
where we used equations (\ref{gmin}) and (\ref{Bsq}). Therefore the Compton parameter
during the electron radiative phase is
\beq
 Y_r = \frac{1}{2} \left(\sqrt{ \frac{5-s}{2(3-s)} \frac{\epsel}{\epsmag}+1 } -1 \right) \;.
\label{Yr}
\eeq
Hence the electron cooling is dominated by IC scatterings (\ie $Y_r > 1$)
for $\epsmag < \eBzero$, where
\beq
 \eBzero = \frac{5-s}{16(3-s)} \epsel \;.
\label{epsB0}
\eeq
Note that $Y_r$ is time-independent. Therefore the $\nu_c$ given by equations (\ref{nucr0}) 
and (\ref{nucr2}) decreases with time for $s=0$ and increases in the $s=2$ model. Thus, 
for observations made at a fixed frequency, the electrons emitting at that frequency change 
their cooling regime from adiabatic to radiative in the case of a homogeneous external gas, 
and from radiative to adiabatic for an external wind (Chevalier \& Li 2000). 

 With the aid of equations (\ref{gmin}), (\ref{Bsq}), and (\ref{trad}), it can be shown that 
the electrons are radiative if
\beq
 n r \Gamma^2 > \frac{3(5-s)}{32(Y_r+1)} \frac{(m_e/m_p)^2}{\sigma_e \epsel \epsmag}
        = \frac{4.2 \times 10^{16}}{\epsel \epsmag} \frac{5-s}{Y_r+1}\;{\rm cm^{-2}}\;,
\label{nrG2}
\eeq
which, with the further use of  equations (\ref{Gms0}) -- (\ref{ns2}), leads to the conclusion
that the electrons are radiative until the observer time $T_r$ given by
\beq
 (s=0) \quad T_r = 0.025 \;\E53 \nex0 (Y_r+1)^2 \epse1^2 \eBtwo^2 \;{\rm day}\;,
\label{Tr0}
\eeq
\beq
 (s=2) \quad T_r = 0.23 \; A_* (Y_r+1) \epse1 \eBtwo \;{\rm day}\;.
\label{Tr2}
\eeq

\subsubsection{Adiabatic Electrons}
\label{adbels}

 For $\gamma_i \ll \gamma_c$ equations (\ref{YC}) and (\ref{Nga}) give the Compton parameter 
\beq
 Y_a = \frac{4}{3}\, \tau_e \times \left\{ \begin{array}{ll} 
        \gamma_i^{p-1} \gamma_c^{3-p} & 2 < p < 3  \\ \gamma_i^2 & 3 < p
         \end{array} \right. \;.
\label{Ya}
\eeq

{\bf Case 1: $2 < p < 3$}. 
By substituting equations (\ref{gmin}) and (\ref{gc}) in equation (\ref{Ya}), one obtains
\beq
 Y_a (Y_a+1)^{3-p} = {\cal F}_p (T) \equiv c_s(p) \epsel^{p-1} 
                                    \epsmag^{p-3} (n \Gamma^2 r_{18})^{p-2} \;,
\label{Ya0}
\eeq
where $\log c_s(p)=(3-p)\log(5-s)-\log(3-s)+1.4p-3.7$. 
The Compton parameter can be obtained by solving numerically the above equation.
For analytical purposes, one can approximate $Y_a = {\cal F}_p$ for ${\cal F}_p < 1$,
in which case the IC losses are less important, and 
\beq
 Y_a = {\cal F}_p^{\frac{1}{4-p}} \quad {\rm for} \quad {\cal F}_p > 1 \;. 
\label{Ya1}
\eeq
In the latter case the IC scatterings affect the electron cooling.

 Note that the quantity $n \Gamma^2 r$ in equation (\ref{Ya0}) decreases with time. 
Thus, for $\epsmag < \eBzero$, the Compton parameter $Y_a$ is above unity until a time 
$T_y$ which can be 
determined by substituting equations (\ref{Gms0}) -- (\ref{ns2}) in (\ref{Ya1}):
\beq
 (s=0) \quad T_y = 10^{\frac{8-3p}{p-2}}\; \E53 \nex0 \epse1^{2\frac{p-1}{p-2}}
                     \eBthree^{-2\frac{3-p}{p-2}}\; {\rm day} \;,
\label{Ty01}
\eeq
\beq
 (s=2) \quad T_y = 10^{\frac{4.9-1.6p}{p-2}}\; A_* \epse1^{\frac{p-1}{p-2}}
                      \eBthree^{-\frac{3-p}{p-2}} \; {\rm day} \;.
\label{Ty21}
\eeq
Thus, for $\epsmag < \eBzero$ and $T_r < T < T_y$, the Compton parameter determines 
the evolution of the cooling break frequency (\eqs [\ref{nucr0}] and [\ref{nucr2}]):
\beq
 \nu_c \stackrel{(s=0)}{=} 10^{15 + \frac{2.5p-5.5}{4-p}} 
    \left[ \E53^{-\frac{p}{2}} \nex0^{-2} \epse1^{-2(p-1)} \eBthree^{-\frac{p}{2}} 
       T_d^{\frac{3p-8}{2}} \right]^{\frac{1}{4-p}}\; {\rm Hz} \;,
\label{nucr01}
\eeq
\beq
 \nu_c \stackrel{(s=2)}{=} 10^{15 + \frac{2.2p-5.5}{4-p}} \E53^{\frac{1}{2}}
                 \left[ A_*^{-4} \epse1^{-2(p-1)} \eBthree^{-\frac{p}{2}} 
                 T_d^{\frac{3p-4}{2}} \right]^{\frac{1}{4-p}} \; {\rm Hz} \;.
\label{nucr21}
\eeq
Note that for a homogeneous medium ($s=0$) and $8/3 < p < 3$, the cooling break 
frequency increases with time, unlike the decreasing behavior it has for $T < T_r$.

{\bf Case 2: $p > 3$}.
This case is treated here for completeness, as there are no afterglows for which such a 
steep electron index has been found. Equations (\ref{gmin}), (\ref{tau}), and (\ref{Ya})
lead to
\beq
 Y_a = \frac{3}{3-s} \epsel^2 n\Gamma^2 r_{18} \;.
\label{Ya2}
\eeq
For $\epsmag < \eBzero$ the Compton parameter is above unity until
\beq
 (s=0) \quad T_y = 0.11\; \E53 \nex0 \epse1^4\; {\rm day} \;,
\label{Ty02}
\eeq
\beq
 (s=2) \quad \quad T_y = 0.87\;  A_* \epse1^2\; {\rm day} \;.
\label{Ty22}
\eeq
For $\epsmag < \eBzero$ and $T_r < T < T_y$, the evolution of the cooling break frequency is
\beq
 (s=0) \quad \nu_c = 1.1 \times 10^{17}\; \E53^{-3/2} \nex0^{-2} \epse1^{-4}
                         \eBthree^{-3/2} T_d^{1/2} \; {\rm Hz} \;,
\label{nucr02}
\eeq
\beq
 (s=2) \quad \nu_c = 1.5 \times 10^{16}\; \E53^{1/2} A_*^{-4} \epse1^{-4}
                         \eBthree^{-3/2} T_d^{5/2} \; {\rm Hz} \;.
\label{nucr22}
\eeq
Note that in this regime $\nu_c$ increases with time in both models.

\subsection{Absorption Break}

 The synchrotron self-absorption frequency $\nu_a$ can be calculated with the aid of 
equation (6.50) from Rybicki \& Lightman (1979). With the notations $\gamma_p = \min 
(\gamma_i,\gamma_c)$, $\nu_p = \min (\nu_i, \nu_c)$, and $\nu_0 = \max (\nu_i, \nu_c)$ 
it can be shown that optical thickness to synchrotron self-absorption can be approximated by
\beq
 \tau_{ab}({\nu}) \simeq 5 \frac{e \Sigma}{B \gamma_p^5} \times 
            \left\{ \begin{array}{ll} (\nu/\nu_p)^{-5/3} & \nu < \nu_p \\
            (\nu/\nu_p)^{-(q+4)/2} & \nu_p < \nu < \nu_0 \end{array} \right.  \;,
\label{tauab}
\eeq
where $\Sigma = (3-s)^{-1} nr$ is the remnant electron column density, $q=2$ for radiative 
electrons ($\gamma_c < \gamma_i$), $q=p$ for adiabatic electrons ($\gamma_i < \gamma_c$).

\subsubsection{Radiative Electrons}

 Equations (\ref{Bsq}), (\ref{gc}), and (\ref{tauab}) give the optical thickness
at the cooling break frequency 
\beq
 \tau_c = \frac{5}{3-s} \frac{enr}{B\gamma_c^5} = \frac{1.1\times 10^{-3}}{(3-s)(5-s)^5} 
          (Y_r+1)^5 \epsmag^{9/2} n^{11/2} \Gamma^4 r_{18}^6 \;.
\label{tauc}
\eeq

% {\bf Homogeneous external medium.} 
 For $s=0$ equations (\ref{Gms0}), (\ref{rs0}), and (\ref{tauc}) lead to
\beq
 (s=0) \quad \tau_c = 0.11\; \E53^2 \nex0^{7/2} (Y_r+1)^5 \epsmag^{9/2} \;.
\label{tauc0}
\eeq
For $\tau_c < 1$ the optical thickness to synchrotron self-absorption is unity at 
$\nu_a$ given by $\nu_a = \nu_c \tau_c^{3/5}$: 
\beq
 (s=0) \; \nu_a = 6.5 \times 10^9\; \E53^{7/10} \nex0^{11/10}
             (Y_r+1) \eBone^{6/5} T_d^{-1/2} \; {\rm Hz} \;.
\label{nuab01}
\eeq

%{\bf Pre-ejected wind.} 
 For $s=2$ equations (\ref{Gms2}) -- (\ref{ns2}), and 
(\ref{tauc}) give
\beq
 (s=2) \qquad \tau_c = 0.44\; \E53^{-3/2} A_*^7 (Y_r+1)^5 \epsmag^{9/2} T_d^{-7/2} \;.
\label{tauc2}
\eeq
For $T > T_a$ we have $\tau_c < 1$ and $\nu_a < \nu_c$:
\beq
 (s=2) \quad \nu_a = 1.4 \times 10^{12}\; \E53^{-2/5} A_*^{11/5}
                  (Y_r+1) \eBtwo^{6/5} \Tdtwo^{-8/5} \; {\rm Hz} \;.
\label{nuab22}
\eeq

\subsubsection{Adiabatic Electrons}

 The optical thickness to synchrotron self-absorption at the injection break can
be found using equations (\ref{gmin}), (\ref{Bsq}), and (\ref{tauab}):
\beq
 \tau_i = \frac{5}{3-s} \frac{enr}{B\gamma_i^5} = \frac{2.9\times 10^{-2}}{3-s} 
           \frac{n^{1/2} r_{18}}{\epse1^5 \epsmag^{1/2} \Gamma^6} \;.
\label{taum}
\eeq

 If the external medium is homogeneous
\beq
 (s=0) \quad \tau_i = 1.3 \times 10^{-6}\; \E53^{-1/2} \nex0 \epse1^{-5}
                      \eBtwo^{-1/2} T_d^{5/2}  \;,
\label{taum0}
\eeq 
\beq
 (s=0) \qquad \nu_a = 2.6 \times 10^9\; \E53^{1/5} \nex0^{3/5} \epse1^{-1} 
                     \eBtwo^{1/5} \; {\rm Hz} \;,
\label{nuab03}
\eeq
which is time-independent.

 For the wind model
\beq
 (s=2) \quad \tau_i = 6.8 \times 10^{-7}\; \E53^{-3/2} A_*^2 \epse1^{-5}
                          \eBtwo^{-1/2} T_d^{3/2}  \;,
\label{taum2}
\eeq 
\beq
 (s=2) \quad \nu_a = 3.7 \times 10^9\; \E53^{-2/5} A_*^{6/5} \epse1^{-1}
                     \eBtwo^{1/5} T_d^{-3/5}\; {\rm Hz} \;.
\label{nuab23}
\eeq 

Note that, in general, the absorption frequency decreases faster for a remnant 
interacting with a wind than for one running into a homogeneous external medium.

\section{Analytical Light-Curves}

 If the effects arising from the remnant spherical shape (see Appendix) are ignored,
than the observed flux peaks at $\nu_p = \min (\nu_i, \nu_c)$, where it has a value 
\beq
  F_{\nu_p}\simeq \frac{\sqrt{3} \phi_p}{4\pi D^2} \frac{e^3}{m_e c^2} \Gamma B N_e \;.
\label{Fnusy}
\eeq
Here $\phi_p$ is a factor calculated by Wijers \& Galama (1999), which we shall set 
$\phi_p=0.63$, $D = (1+z)^{-1/2} D_l(z)$ with $D_l$ the luminosity distance, and 
$N_e = m(r)/m_p$ is the number of electrons in the remnant. Equations (\ref{gmin}), 
(\ref{Bsq}), and (\ref{Fnusy}) give
\beq
 F_{\nu_p} = \frac{57}{3-s}\D28 \epsmag^{1/2} \Gamma^2 n^{3/2} r_{18}^3\;{\rm mJy}\;. 
\label{Fnupeak}
\eeq

 The afterglow emission at any given frequency and time can be calculated using 
the synchrotron spectrum for the electron distributions given in equations (\ref{Ngr})
and (\ref{Nga}) (\eg Sari \etal 1998):
\beq
 F_{\nu} = F_{\nu_c} \left\{ \begin{array}{lll}
  (\nu/\nu_a)^2 (\nu_a/\nu_c)^{1/3}            &  \nu < \nu_a  &{\rm (1)}\\
  (\nu/\nu_c)^{1/3}                            &\nu_a<\nu<\nu_c&{\rm (2)}\\
  (\nu/\nu_c)^{-1/2}                           &\nu_c<\nu<\nu_i&{\rm (3)}\\
  (\nu/\nu_i)^{-p/2}(\nu_c/\nu_i)^{1/2}        &  \nu_i < \nu  &{\rm (4)}
                             \end{array} \right.  
\label{Fnur}
\eeq 
for $T < T_r$, assuming that $\nu_a < \nu_c$, and
\beq
 F_{\nu} = F_{\nu_i} \left\{ \begin{array}{lll}
  (\nu/\nu_a)^2 (\nu_a/\nu_i)^{1/3}            &  \nu < \nu_a  &{\rm (5)}\\
  (\nu/\nu_i)^{1/3}                            &\nu_a<\nu<\nu_i&{\rm (6)}\\
  (\nu/\nu_i)^{-(p-1)/2}                       &\nu_i<\nu<\nu_c&{\rm (7)}\\
  (\nu/\nu_c)^{-p/2} (\nu_i/\nu_c)^{(p-1)/2}   &  \nu_c < \nu  &{\rm (8)}
                                             \end{array} \right.  
\label{Fnua}
\eeq 
for $T > T_r$, assuming that $\nu_a < \nu_i$.

 In Figures 1 and 2 the plane $T-\epsmag$ is divided into several regions which
are labeled as in equations (\ref{Fnur}) -- (\ref{Fnua}), according to the ordering 
of the observing frequency $\nu$ and of the three break frequencies $\nu_a$, $\nu_i$, 
and $\nu_c$. The observed fluxes in each case are given in the Appendix, and a set of 
multi-wavelength light-curves is shown in Figure 3. The results shown in Figures 1--4
have been obtained using equations which are valid in any relativistic regime, such
as equations (\ref{Gm}), (\ref{gmin}), (\ref{Bsq}), and (\ref{Fnusy}).

\subsection{Inverse Compton Emission}

 The IC emission can be easily calculated by using the above equations
for the synchrotron spectrum and Compton parameter. The up-scattered spectrum peaks
at $\nu_{ic} \sim \gamma_c^2 \nu_c$ if electrons are radiative and at $\nu_{ic} \sim
\gamma_i^2 \nu_i$ if electrons are adiabatic. It can be shown that, for any electron 
radiative regime, the flux of the up-scattered emission at this frequency is
\beq
 F^{(ic)}_{\nu_{ic}} = \tau_e F_{\nu_p} \;,
\label{Fic}
\end{equation}
where $\tau_e$ and $F_{\nu_p}$ are given by equations (\ref{tau}) and (\ref{Fnupeak}),
respectively. For the up-scattering of synchrotron photons above $\nu_a$ and assuming
that $Y_a < 1$, the resulting IC light-curves have the following behaviors:
\beq
 \left( \begin{array}{l}  s=0 \\ T < T_r \end{array} \right) 
              \quad  F^{(ic)}_{\nu} \propto \left\{ \begin{array}{ll} 
                       T^{1/3} & \nu \siml \gamma_c^2 \nu_c \\
                       T^{1/8} & \gamma_c^2 \nu_c \siml \nu \siml \gamma_i^2 \nu_i \\
                       T^{-(9p-10)/8} & \gamma_i^2 \nu_i \siml \nu 
                    \end{array}       \right. \;,
\label{Fic0r}
\eeq
\beq
 \left( \begin{array}{l}  s=0 \\ T > T_r \end{array} \right) 
       \quad  F^{(ic)}_{\nu} \propto \left\{ \begin{array}{ll} 
                       T^{1} & \nu \siml \gamma_i^2 \nu_i \\
                       T^{-(9p-11)/8} & \gamma_i^2 \nu_i \siml \nu \siml \gamma_c^2 \nu_c \\
                       T^{-(9p-10)/8} & \gamma_c^2 \nu_c \siml \nu 
                   \end{array}      \right. \;,
\label{Fic0a}
\eeq
\beq
 \left( \begin{array}{l}  s=2 \\ T < T_r \end{array} \right) 
          \quad  F^{(ic)}_{\nu} \propto \left\{ \begin{array}{ll} 
                       T^{-5/3} & \nu \siml \gamma_c^2 \nu_c \\
                       T^0 & \gamma_c^2 \nu_c \siml \nu \siml \gamma_i^2 \nu_i \\
                       T^{-(p-1)} & \gamma_i^2 \nu_i \siml \nu 
                    \end{array}       \right. \;,
\label{Fic2r}
\eeq
\beq
 \left( \begin{array}{l}  s=2 \\ T > T_r \end{array} \right) 
         \quad  F^{(ic)}_{\nu} \propto \left\{ \begin{array}{ll} 
                       T^{-1/3} & \nu \siml \gamma_i^2 \nu_i \\
                       T^{-p} & \gamma_i^2 \nu_i \siml \nu \siml \gamma_c^2 \nu_c \\
                       T^{-(p-1)} & \gamma_c^2 \nu_c \siml \nu 
                 \end{array}      \right. \;.
\label{Fic2a}
\eeq

 For external media that are not denser than assumed so far, the IC emission is weaker 
than synchrotron, even in soft X-rays. As shown in the upper left panel of Figure 3, 
for $n_* \simg 10\;{\rm cm^{-3}}$ and $A_* \simg 1$ the up-scattered radiation can 
dominate the synchrotron emission at times $T \simg 10^{-1}$ day, diminishing the decay 
rate of the X-ray emission. This is due to that the flux at the IC peak (\eq [\ref{Fic}]) 
depends strongly on the external medium density: $F^{(ic)}_{\nu_{ic}} \propto n_*^{5/4}$ 
for $s=0$ and $F^{(ic)}_{\nu_{ic}} \propto A_*^{5/2}$ for $s=2$. The afterglow flattening 
is strongly dependent on the observing frequency, being absent in the optical and below.

\section{Conclusions}

 Using the analytical results given in equations (\ref{F01}) -- (\ref{F28}), the 
afterglow light-curve can be calculated at any frequency and at observing times
up to the onset of the non-relativistic phase. As illustrated in Figure 3, the largest 
differences between the afterglow emission in the two models for the external medium 
is seen at low frequencies (lower panels). However, the scintillation due to the local 
interstellar medium (Goodman 1997), may hamper the use of the radio light-curves to 
identify the type of external medium and geometry of the ejecta (Frail \etal 2000). 

 Figure 4 shows that, for various model parameters, the rate of change of the afterglow 
emission at $\nu \sim 10^{12}$ Hz and at early times (when the jet effects are negligible,
provided that the jet is initially wider than a few degrees) exhibits a strong dependence 
on the type of external medium. If the external medium is homogeneous
the sub-millimeter afterglow should rise slowly at times between $\sim 1$ hour and 
$\sim 1$ day, while for a pre-ejected wind the emission should fall off steeply, 
followed by a plateau\footnotemark . 
\footnotetext{The light-curves presented in the upper left panel of Figure 4 
 show that this criterion for determining the type of external medium fails only if the 
 particle density of the homogeneous medium exceeds $\sim 10\;{\rm cm^{-3}}$. In this 
 case the X-ray emission may help to distinguish between the two models of external 
 medium, as the absence of a flattening of the high energy emission is compatible
 only with a pre-ejected wind with $A_*$ less than a few.}
Therefore observations made at sub-millimeter frequencies with the SCUBA (James Clerk 
Maxwell Telescope) or with the MAMBO (IRAM Telescope) instruments would be very powerful 
in determining if the medium which the remnant runs into is homogeneous or follows a 
$r^{-2}$ law. 

 We note that if turbulence in the shocked fluid does not lead to a significant
mixing, then the inhomogeneous electron distribution will alter the afterglow spectrum 
below the absorption frequency $\nu_a$ as described by Granot, Piran \& Sari (2000). 
The result is that the afterglow emission at $\nu < \nu_a$ rises more slowly than calculated 
here. For instance, the $T^2$ rise exhibited by the $\nu = 10^{12}$ Hz light-curves shown 
in Figure 4 for the wind model at early times becomes a $T^1$ rise. Nevertheless, the 
basic difference mentioned above between the temporal behaviors of the sub-millimeter 
light-curves at $10^{-2}\,{\rm day} \siml T \siml 1\,{\rm day}$ remains unchanged.

 The IC losses alter the evolution of the cooling break $\nu_c$ if the 
electrons injected with minimal energy are adiabatic and if the Compton parameter is above 
unity (\ie the magnetic field parameter is weaker than that given in \eq [\ref{epsB0}]). 
In the case of a homogeneous external medium, the cooling break
frequency decreases as $T^{-1/2}$ if the electrons are radiative. When the electrons 
become adiabatic, this break evolves as $T^{\frac{3p-8}{8-2p}}$ for $p < 3$ and increases 
as $T^{1/2}$ for $p > 3$. For an external wind the change is from $\nu_c \propto T^{1/2}$ 
to $\nu_c \propto T^{\frac{3p-4}{8-2p}}$ for $p < 3$, and to $\nu_c \propto T^{5/2}$
for $p > 3$. Consequently the power-law decay of the afterglow emission at frequencies 
above the cooling break flattens by up to $1/2$ if the external medium is homogeneous and 
by up to $1$ if the medium is a wind. For an electron index $p < 3$, the flattening is 
mild and likely to be seen only in the optical emission from a remnant interacting with 
a pre-ejected wind. 

 The IC emission itself is generally weaker than the synchrotron emission. 
Nevertheless, if the external medium is sufficiently dense (\ie $n_* \simg 10\,{\rm cm^{-3}}$
or $A_* \simg$ few), a flattening of the soft X-ray light-curve should be seen few hours 
after the main event, at fluxes well above the threshold of BeppoSAX (see Figure 3, upper 
left panel). The flattening of the afterglow emission due to the up-scattered radiation is 
a chromatic feature, appearing only at high frequencies, and its strength is moderately 
dependent on the remaining model parameters. 

 Finally, another possible signature of the interaction with the wind ejected by a 
Wolf-Rayet star should be found during the GRB emission in the form of smooth pulses of 
increasing duration. Such pulses are generated in internal shocks when the decelerating 
outermost shell is hit from behind by shells ejected at later times. This phenomenon
is more likely to be seen in the wind model, for which the deceleration radius is
smaller than for a homogeneous medium.

\acknowledgements{AP acknowledges the support received from Princeton University through
 the Lyman Spitzer, Jr. Fellowship. We thank Bohdan Paczy\'nski for useful discussions.}

\begin{appendix}

\vspace*{2mm}

 The observer time calculated with equation (\ref{dT}) is the time when photons 
emitted by the shocked gas moving precisely toward the observer arrive at detector. 
Photons emitted by the fluid moving at a non-zero angle relative to the direction
toward the observer are less boosted by the relativistic motion of the source 
and arrive at the observer later. The effect of the remnant spherical curvature
on the afterglow emission can be reduced to a correction factor that must be applied 
to the light-curve obtained analytically using the remnant parameters at laboratory
frame time $t$ related to the observer time $T$ through equation (\ref{dT}). 
This correction factor is dependent on the observing frequency, as described below.

\vspace*{3mm}
\section{Correction Factors for Analytical Light-Curves}

 The flux received by the observer at time $T$ can be calculated by integrating the remnant
emission over the equal arrival-time surface (Panaitescu \& \Mesz 1998). This surface is
defined by $cT=ct-r\cos\theta$, where $\theta$ is the azimuthal angle measured relative to 
the observer's line of sight toward the remnant center. Using the equation for dynamics
of relativistic remnants ($\Gamma \propto r^{-(3-s)/2}$), the integral can be written as
\beq
 F_\nu(T) = \frac{A (cT)^{-2}}{2(3-s) D^2} \int_0^{R(T)} 
            \frac{ (P'_{\nu'})_e r^{4-s} }{\Gamma^3 (x+1)^2} {\rm d}r \;,
\label{Fnu1}
\eeq
where $x \equiv (3-s)(r/R)^{4-s}$, and $R$ is the radius for which photons emitted 
at $\theta=0$ arrive at $T$, \ie the maximal radius on the equal arrival-time surface. 
In equation (\ref{Fnu1}) $(P'_{\nu'})_e$ is the co-moving frame power per electron, 
taking into account the spectrum of the emission. For the synchrotron spectra given in 
equations (\ref{Fnur}) and (\ref{Fnua}), one can write generically $(P'_{\nu'})_e = 
\sqrt{3} \phi_p (e^3 B/m_e c^2)(\nu'/\nu'_a)^{\alpha_a} (\nu'/\nu'_i)^{\alpha_i} 
(\nu'/\nu'_c)^{\alpha_c}$, where $\nu'$ is the co-moving frame observing frequency.
The integral in equation (\ref{Fnu1}) is determined by the the values at of the integrand
at $r \siml R$.
 
 If the effect of the remnant geometrical curvature on the photon arrival-time were 
ignored, then the observed flux would be 
\beq
 F_\nu^{(0)} (R) = \frac{1}{4 \pi D^2} \left[ (P'_{\nu'})_e \Gamma N_e \right]_{R(T)} \;,
\label{Fnu2}
\eeq
which can be calculated based on equations (\ref{Fnusy}), (\ref{Fnur}) and (\ref{Fnua}).
The $r$-dependent quantities in equation (\ref{Fnu1}) can be expressed in terms of their
values at $r=R$, so that the flux $F_\nu$ can be written as $F_\nu^{(0)}$ times a correction 
factor 
\beq
 K = 2\,(4-s)^{2-(\alpha_a+\alpha_i+\alpha_c)} \int_0^1 u^{f(s)} 
          \left[ 1+(3-s)u^{4-s} \right]^{\alpha_a+\alpha_i+\alpha_c-2} {\rm d}u \;,
\label{K}
\eeq
where $f(s) = 7 - 2.5s - 2\alpha_a(q-3s) + \alpha_i(2-0.5s) - \alpha_c(2+0.5s)$,
with $q=1,2$ for radiative, radiative electrons, respectively. The $K$ factor depends
on $\nu$ and also on $p$ if $\nu > \nu_i$. Table 1 gives its values for various cases.

\vspace*{3mm}
\section{Homogeneous External Medium ($s=0$)} 

 By substituting the equations for the break frequencies and equation (\ref{Fnupeak})
in equations (\ref{Fnur}) and (\ref{Fnua}), and taking into account the above correction 
factors for the remnant curvature, the following fluxes are obtained
\beq
 F_{\nu} \stackrel{(1)}{=} 0.3\; \D28 (Y_r+1)^{-1} \nex0^{-1} \eBone^{-1}\; 
                           \nu_{9.7}^2 \Tdone \; {\rm mJy} \;,
\label{F01}
\eeq
\beq
 F_{\nu} \stackrel{(2)}{=} 10\; \D28 (Y_r+1)^{2/3} \E53^{7/6} \nex0^{5/6} \eBtwo\;
                           \nuo^{1/3} \Tdtwo^{1/6} \; {\rm mJy} \;,
\eeq
\beq
 F_{\nu} \stackrel{(3)}{=} 40\; \D28 (Y_r+1)^{-1} \E53^{3/4}  \eBone^{-1/4}\;
                           \nuo^{-1/2} \Tdtwo^{-1/4} \; {\rm mJy} \;,
\eeq
\beq
 F_{\nu} \stackrel{(4)}{=} 10^{2.1-0.6p}\; \D28 (Y_r+1)^{-1} \E53^{\frac{p+2}{4}} \epse1^{p-1} 
         \eBone^{\frac{p-2}{4}}\; \nuo^{-\frac{p}{2}} T_d^{-\frac{3p-2}{4}} \; {\rm mJy} \;,
\eeq
\beq
 F_{\nu} \stackrel{(5)}{=} 30\; \D28 \E53^{1/2} \nex0^{-1/2} \epse1 \; 
                           \nu_{9.7}^2 T_{d,1}^{1/2} \; {\rm mJy} \;,
\eeq
\beq
 F_{\nu} \stackrel{(6)}{=} 1\; \D28 \E53^{5/6} \nex0^{1/2} \epse1^{-2/3} 
                         \eBfour^{1/3}\; \nuo^{1/3} \Tdtwo^{1/2} \; {\rm mJy} \;,
\eeq
\beq
 F_{\nu} \stackrel{(7)}{=} 10^{2.1-1.3p}\; \D28 \E53^{\frac{p+3}{4}} \nex0^{1/2} 
                           \epse1^{p-1} \eBfour^{\frac{p+1}{4}}\; 
                   \nuo^{-\frac{p-1}{2}} T_d^{-\frac{3}{4}(p-1)} \; {\rm mJy} \;,
\eeq
\beq
 F_{\nu} \stackrel{(8)}{=} 10^{2.4-0.8p}\; \D28 \E53^{\frac{p+2}{4}} \epse1^{p-1} 
   \eBtwo^{\frac{p-2}{4}}\; \nuo^{-\frac{p}{2}} T_d^{-\frac{3p-2}{4}} \; {\rm mJy} \;,
\eeq
\beq
 F_{\nu} \stackrel{(8a)}{=} 10^{\frac{2p^2-7.7p+0.8}{4-p}}\; 
     \D28 \left[ \E53^{\frac{1}{4}(12-p^2)} \nex0^{-\frac{1}{2}(p-2)} \epse1^{(p-1)(3-p)} 
         \eBfour^{\frac{1}{4}(-p^2+2p+4)} \right]^{\frac{1}{4-p}}\; \nu_{17.5}^{-\frac{p}{2}} 
            \Tdone^{-\frac{3p}{4}+\frac{1}{4-p}}\; {\rm mJy} \quad (2 < p < 3)\;.
\label{F08a}
\eeq

 The case given in equation (\ref{F08a}) and labeled $(8a)$ corresponds to the same 
frequency ordering as for case $(8)$, but the cooling break $\nu_c$ evolution 
(\eq [\ref{nucr01}]) is determined by the IC losses, \ie $T_r < T < T_y$ 
and $Y_a > 1$.

\vspace*{3mm}
\section{Wind External Medium ($s=2$)}

 Following the same exercise as above and using the relevant equations, the following 
results can be obtained for the wind model:
\beq
 F_{\nu} \stackrel{(1)}{=} 0.03\; \D28 (Y_r+1)^{-1} \E53 A_*^{-2} \epse1^{-1}\; 
                           \eBtwo^{-1} \nu_{9.7}^2 \Tdone^2 \; {\rm mJy} \;,
\eeq
\beq
 F_{\nu} \stackrel{(2)}{=} 70\; \D28 (Y_r+1)^{2/3} \E53^{1/3} A_*^{5/3} \eBtwo\;
                           \nu_{12}^{1/3} \Tdone^{-2/3} \; {\rm mJy} \;,
\eeq
\beq
 F_{\nu} \stackrel{(5)}{=} 0.07\; \D28 \E53 A_*^{-1} \epse1 \; 
                           \nu_{9.7}^2 \Tdone^1 \; {\rm mJy} \;,
\eeq
\beq
 F_{\nu} \stackrel{(6)}{=} 9\; \D28 \E53^{1/3} A_* \epse1^{-2/3} 
                       \eBthree^{1/3}\; \nu_{12}^{1/3} T_d^0\; {\rm mJy} \;,
\eeq
\beq
 F_{\nu} \stackrel{(7)}{=} 10^{2.3-1.2p}\; \D28 \E53^{\frac{p+1}{4}} A_* \epse1^{p-1} 
  \eBfour^{\frac{p+1}{4}}\; \nuo^{-\frac{p-1}{2}} T_d^{-\frac{3p-1}{4}} \; {\rm mJy} \;,
\eeq
\beq
 F_{\nu} \stackrel{(8a)}{=} 10^{\frac{1.9p^2-8.6p+5.4}{4-p}}\; 
       \D28 \E53^{\frac{p+2}{4}} \left[ A_*^{-(p-2)} \epse1^{(p-1)(3-p)} 
      \eBfour^{\frac{1}{4}(-p^2+2p+4)} \right]^{\frac{1}{4-p}}\; \nu_{17.5}^{-\frac{p}{2}} 
      \Tdone^{-\frac{3p}{4}+\frac{p}{2(4-p)}}\; {\rm mJy} \quad (2 < p < 3)\;.
\eeq

 For the remainder of the cases it can be shown that the fluxes for the wind
model differ from those for obtained in the $s=0$ model only by a constant factor:
\beq
 \frac{F_{\nu}(s=2)}{F_{\nu}(s=0)} \stackrel{(3)}{=} 0.60\; \frac{Y_r(s=0)}{Y_r(s=2)} \;,
\eeq
\beq
 \frac{F_{\nu}(s=2)}{F_{\nu}(s=0)} \stackrel{(4)}{=} 
       1.24\; \left( \frac{17}{72} \right)^{p/4} \frac{Y_r(s=0)}{Y_r(s=2)} \;,
\eeq
\beq
 \frac{F_{\nu}(s=2)}{F_{\nu}(s=0)} \stackrel{(8)}{=} 1.35\; \left( \frac{17}{72} \right)^{p/4} \;.
\label{F28}
\eeq
This means that observations made at frequencies above the cooling break are unable
to distinguish between the two models for the external medium, with the exception
of the case where the electron cooling is dominated by IC losses (case $8a$).

\vspace*{3mm}
 The reference frequency in the equations above was chosen $\nu = 4\times 10^{14}$ Hz 
whenever this optical frequency falls in one of the cases labeled in equations 
(\ref{Fnur})--(\ref{Fnua}), for the model parameters given in Figures 1 and 2. 
The observer time was scaled to a value at which each case is more likely to occur.

 The above light-curves must be corrected to account for cosmological effects. 
This is achieved by replacing $T$ with $T/(1+z)$ and $\nu$ with $(1+z)\nu$, where
$z$ is the burst redshift.

\end{appendix}

\begin{table}[hb]
\begin{center}
 TABLE 1. \\ {\sc Values of the factor $K$ given in equation (\ref{K}), representing the
              correction that should be applied to the analytical light-curves to account
              for the effects due to the remnant geometrical curvature. } \\ [4ex]
 \begin{tabular}{ccccccccc} \hline
 \rule[-4mm]{0mm}{10mm}  $s$  & $(1)$& $(2)$& $(3)$&$(4)^*$& $(5)$& $(6)$&$(7)^*$&$(8)^*$ \\ \hline
 \rule[-2mm]{0mm}{6mm}    0   & 0.50 & 0.55 & 0.59 &  1.37 & 1.00 & 0.42 &  1.34 & 1.02 \\
 \rule[-2mm]{0mm}{6mm}    2   & 0.11 & 1.76 & 0.74 &  1.19 & 0.12 & 0.90 &  2.25 & 0.82 \\ \hline
 \end{tabular}
 \begin{displaymath}
  ^*\; {\rm in\;\; this\;\; case\;\;} K {\;\;\rm depends\;\; on\;\;} p.\;\; 
       {\rm The\;\; values\;\; given\;\; here\;\; are\;\; for\;\;} p=2.5\; .
 \end{displaymath}
\end{center}
\end{table}

\clearpage

\begin{figure*}[tb]
\centerline{\psfig{figure=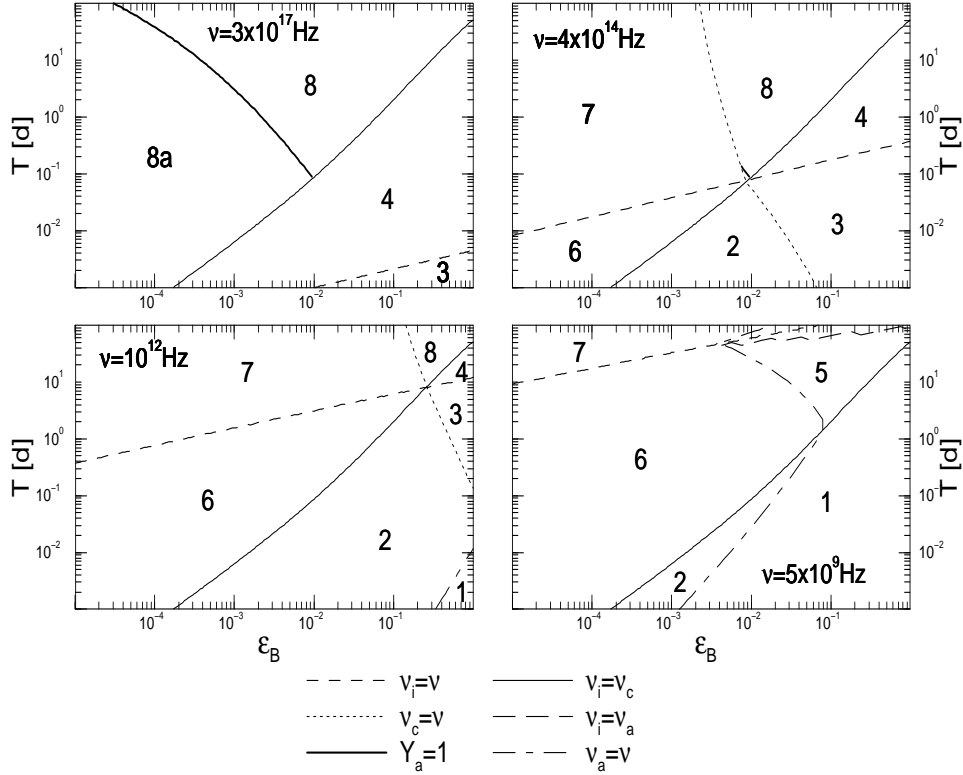,height=10cm,width=13cm}}
\figcaption{The afterglow brightness at a given frequency (indicated in each panel)
 is dependent on the relationship between this frequency and those of the injection 
 ($\nu_i$), cooling ($\nu_c$), and absorption ($\nu_a$) breaks. Different combinations
 of these four frequencies are indicated in the $time - \epsmag$ plane for the homogeneous 
 external medium model ($s=0$), according to the labeling given in equations 
 (\ref{Fnur}) -- (\ref{Fnua}). The label $8a$ indicates the case when $\nu_i < \nu_c < 
 \nu$ and $Y_a > 1$, \ie the cooling break $\nu_c$ evolution is determined by the IC 
 losses.  Other model parameters are $E = 10^{53}$ erg (in $4\pi$ sr), 
 $n_*=1\,{\rm cm^{-3}}$, $\epsel = 0.1$, and $p=2.5$. The upper two panels also indicate 
 the time when the Compton parameter $Y_a$ for adiabatic electrons becomes unity.}
\end{figure*}

\begin{figure*}[tb]
\centerline{\psfig{figure=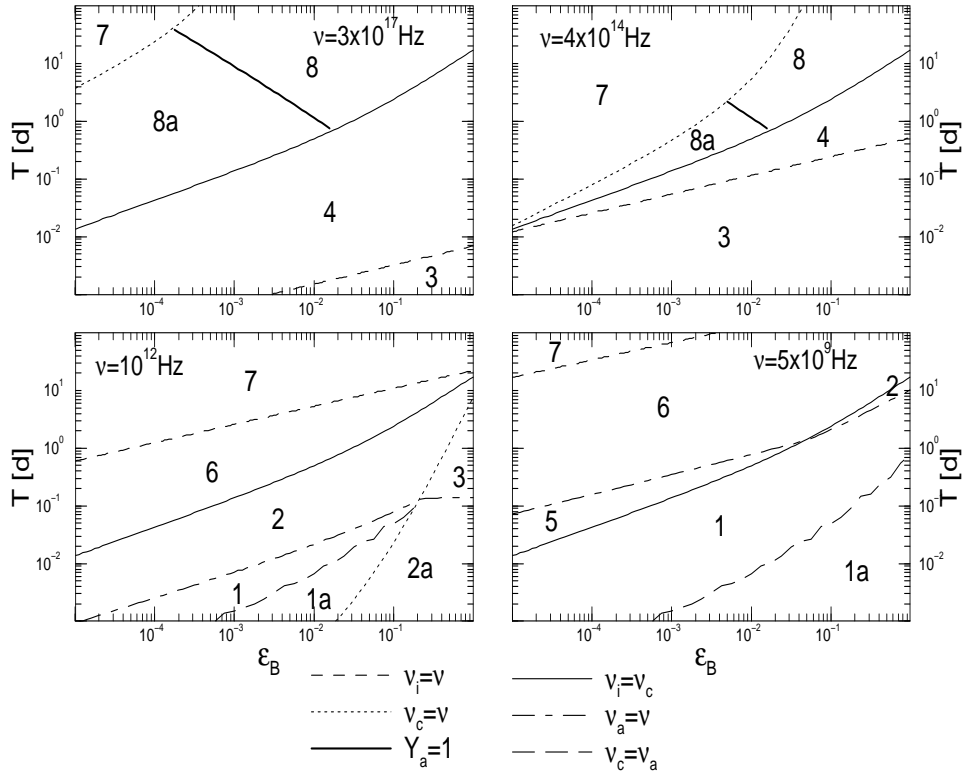,height=10cm,width=13cm}}
\figcaption{Same diagrams as for Figure 1, but for a wind model ($s=2$) with $A_*=1$.}
\end{figure*}

\begin{figure*}[tb]
\vspace*{-1cm}
\centerline{\psfig{figure=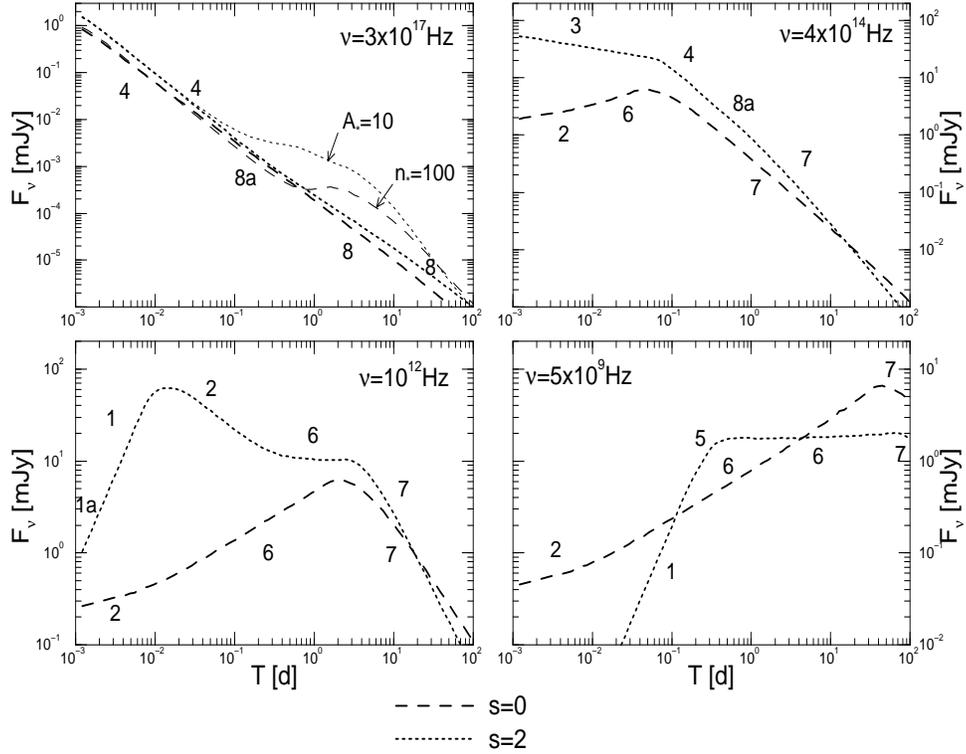,height=10cm,width=13cm}}
\figcaption{Numerical light-curves for $D=10^{28}$ cm, $E = 10^{53}$ erg (in $4\pi$ sr), 
 $\epsel = 0.1$, $\epsmag=10^{-3}$, $p=2.5$, for a homogeneous external medium (dashed 
 lines) with $n_*=1\,{\rm cm^{-3}}$ and a pre-ejected wind (dotted lines) with $A_*=1$, 
 at various observing frequencies (given in each panel). The numbers indicate the cases 
 identified in equations (\ref{Fnur}) - (\ref{Fnua}). The afterglow emission is calculated
 using equations that are valid in any relativistic regime. The effect of the remnant
 geometrical curvature on the photon arrival time and relativistic boosting are taken 
 into account. \newline
 \hspace*{5mm} 
 Note that the largest differences between the two models are shown by the low frequency 
 light-curves at observer times below $\sim 1$ day. \newline
 \hspace*{5mm} 
 Also shown in the upper left panel are two soft X-ray light-curves for denser external 
 media: $n_*=10^2\,{\rm cm^{-3}}$ for $s=0$ (thin dashed curve) and $A_*=10$ for $s=2$ 
 (thin dotted line), in which cases the afterglow emission at $T \simg 10^{-1}$ day is 
 dominated by IC scatterings, and exhibits a substantial flattening. 
 Such a feature is chromatic and cannot be seen at lower observing frequencies. 
 The flattening of the X-ray emission in the case of denser media is present for other
 plausible sets of model parameters, as long as they lead to fluxes (at the time 
 when the flattening occurs) that are detectable with current instruments.}
\end{figure*}

\begin{figure*}[tb]
\centerline{\psfig{figure=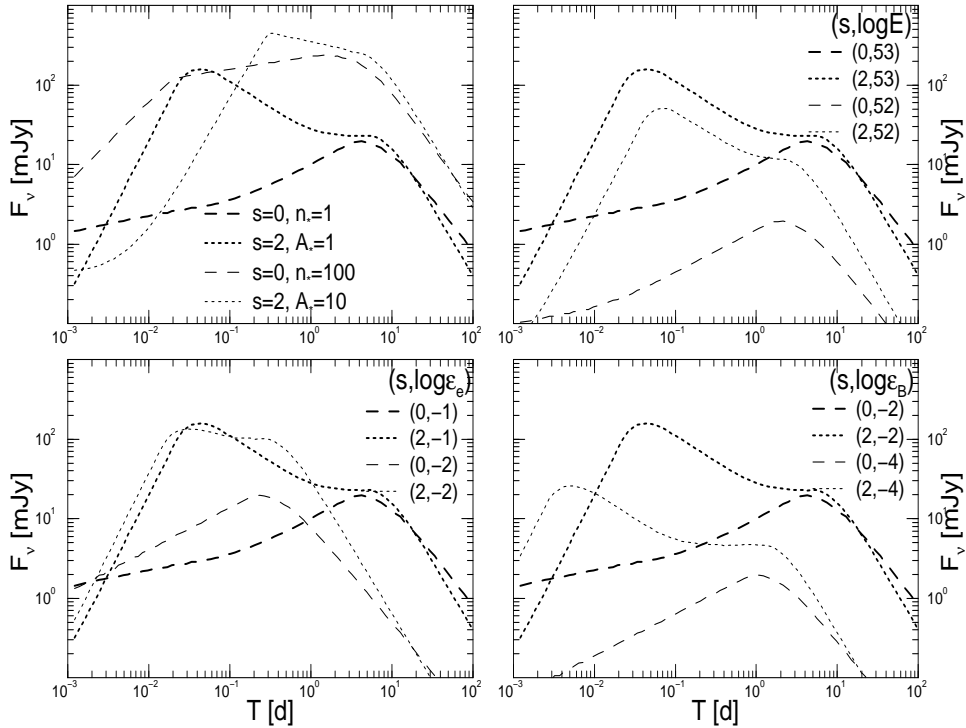,height=10cm,width=13cm}}
\figcaption{Comparison between the afterglow light-curves at observing frequency 
 $\nu = 10^{12}$ Hz for homogeneous external media (dashed lines) and pre-ejected winds 
 (dotted lines), and for various model parameters, as given in each panel. 
 Unless specified in the legend, the model parameters are $E = 10^{53}$ erg 
 (in $4\pi$ sr), $n_*=1\,{\rm cm^{-3}}$, $A_*=1$, $\epsel = 0.1$, $\epsmag=10^{-2}$, 
 and $p=2.5$.}
\end{figure*}

\end{document}